\documentclass[preprint,5p,times,twocolumn]{my-elsarticle}

\usepackage{tensor}
\usepackage{amssymb}
\usepackage{amsmath}
\usepackage{mathrsfs}

\biboptions{compress}

\usepackage[normalem]{ulem}
\usepackage[usenames,table]{xcolor}
\usepackage[colorlinks=true,linkcolor=black,citecolor=blue,urlcolor=blue,filecolor=black]{hyperref}

\usepackage{color}

\newcommand{\cD}{{\cal D}}

\newcommand{\cL}{{\cal L}}

\newcommand{\cO}{{\cal O}}

\newcommand{\cU}{{\cal U}}

\newcommand{\be}{\begin{equation}}
\newcommand{\ee}{\end{equation}}

\newcommand{\pr}{\partial}

\newcommand{\nn}{\nonumber}

\begin{document}

\begin{frontmatter}



\title{Finite actions and asymptotic charges at null infinity for any spin}

\author[UMONS]{Andrea~Campoleoni}
\author[Urbana]{Arnaud Delfante}
\author[RomaTre]{Dario Francia}
\author[IPhT]{Carlo Heissenberg}

\address[UMONS]{Service de Physique de l'Univers, Champs et Gravitation, Universit\'e de Mons -- UMONS, 20 place du Parc, 7000 Mons, 
Belgium\\
\href{mailto:andrea.campoleoni@umons.ac.be}{andrea.campoleoni@umons.ac.be}}
\address[Urbana]{Department of Physics, University of Illinois, 1110 W. Green St., Urbana IL 61801-3080, U.S.A.\\
\href{mailto:delfante@illinois.edu}{delfante@illinois.edu}}
\address[RomaTre]{Roma Tre University and INFN Sezione di Roma Tre, via della Vasca Navale, 84 I-00146 Roma,
Italy\\
\href{mailto:dario.francia@uniroma3.it}{dario.francia@uniroma3.it}}
\address[IPhT]{Institut de Physique Th\'eorique, CEA Saclay, CNRS,
Universit\'e Paris-Saclay, F-91191, Gif-sur-Yvette Cedex, France\\
\href{mailto:carlo.heissenberg@ipht.fr}{carlo.heissenberg@ipht.fr}}

\begin{abstract}
We identify boundary terms renormalizing the free on-shell actions for massless fields of arbitrary spin, including electromagnetism and linearized gravity, with boundary conditions allowing for supertranslation-like asymptotic symmetries. Our focus is on null infinity, in any spacetime dimensions. We also comment on the renormalization of the corresponding asymptotic charges.
\end{abstract}





\end{frontmatter}


\section{Introduction}

The relations between asymptotic symmetries, soft theorems and memory effects established in gravity and gauge theories (see \cite{Strominger:2017zoo} for a review) have stimulated the investigation of new boundary conditions allowing for larger asymptotic symmetries. Extensions of this kind are often motivated by the will to interpret a known infrared effect as the manifestation of an asymptotic symmetry, but one typically has to handle divergences induced by the weaker boundary conditions that one has to impose. A prototypical example is given by $\text{\textit{Diff}}(S^2)$ superrotations, that have been proposed in \cite{Campiglia:2014yka} to account for subleading soft theorems in four-dimensional gravity. The associated surface charges ---~computed using, \emph{e.g.}, the Wald--Zoupas prescription \cite{Wald:1999wa}~--- diverge at null infinity, but they have been renormalized profiting from the intrinsic ambiguities present in their definition~\cite{Compere:2018ylh}. This procedure to renormalize surface charges has been extensively studied in this and related contexts, see \emph{e.g.}  \cite{Papadimitriou:2005ii, Compere:2008us, Freidel:2019ohg, Flanagan:2019vbl, Giddings:2019ofz, Campiglia:2020qvc, Compere:2020lrt, Fiorucci:2020xto, Chandrasekaran:2021vyu, Geiller:2022vto, Campoleoni:2022wmf, Capone:2023roc, McNees:2023tus, Campoleoni:2023eqp, Ciambelli:2023ott, Agrawal:2023zea, Riello:2024uvs, Choi:2024ygx, McNees:2024iyu, Romoli:2024hlc, Francia:2024hja, Choi:2024ajz,Choi:2024mac, Manzoni:2025wyr}  and \cite{Fiorucci:2021pha, Capone:2021aas, Delfante:2024npo} for reviews. While it is by now clear that a prescription giving finite surface charges always exists (see, \emph{e.g.}, \cite{McNees:2023tus, Riello:2024uvs}), its compatibility with further requirements, like covariance or locality, is still under debate \cite{Flanagan:2019vbl, Odak:2022ndm, Capone:2023roc, Rignon-Bret:2024wlu}. Moreover, other quantities, like the on-shell action, might diverge as well.
 
Other notable asymptotic symmetries that require renormalization are supertranslations in asymptotically flat spacetimes of dimension $D>4$. These symmetries have been long neglected since, differently from $D=4$, one can define a radiative solution space with boundary conditions not allowing for them \cite{Hollands:2003ie, Tanabe:2011es, Hollands:2016oma}. On the other hand, even when $D > 4$, supertranslations and their spin-one analogs have been related to soft theorems and memory effects in gravity \cite{Kapec:2015vwa, Mao:2017wvx, Pate:2017fgt, Aggarwal:2018ilg} and Yang--Mills theories \cite{Kapec:2014zla, He:2019jjk, He:2019pll, Campoleoni:2019ptc, He:2023bvv}. This link also holds for fields of arbitrary spin \cite{Campoleoni:2017mbt, Francia:2018jtb, Campoleoni:2020ejn}. Their surface charges, computed according to standard prescriptions, however diverge when $D>4$ \cite{Campoleoni:2020ejn}. On top of that, the boundary conditions allowing for these symmetries also lead to a divergent on-shell action. As we shall discuss, the only exception is given by the on-shell Maxwell action, which, in our setup, is manifestly finite at null infinity in any $D$.

In this Letter, we focus on the renormalization of the on-shell actions for free massless fields of arbitrary spin at null infinity, for the boundary conditions allowing for supertranslation-like asymptotic symmetries identified in \cite{Campoleoni:2017mbt,  Campoleoni:2020ejn}. To this end, we follow the approach of holographic renormalization \cite{deHaro:2000vlm, Bianchi:2001kw, Skenderis:2002wp, Papadimitriou:2016yit}. This has been mainly developed on asymptotically Anti de Sitter (AdS) spacetimes, although some studies on how to regulate on-shell actions in asymptotically flat spacetimes appeared in \cite{Mann:2005yr, Afshar:2018apx, Hartong:2025jpp}. The links between the renormalization of the on-shell action and of the surface charges have also been explored in AdS, see, \emph{e.g.}, \cite{Papadimitriou:2005ii, Compere:2008us, Compere:2020lrt, Fiorucci:2020xto, Campoleoni:2023eqp}, and we shall comment on them in our setup. Aside from this application, our findings are a first step towards a systematic treatment of holographic renormalization in flat space, which might open a new avenue towards flat space holography (see, \emph{e.g.}, \cite{Donnay:2023mrd} for a review).

Concretely, we propose a boundary counterterm canceling the divergences at null infinity of the on-shell action for free massless fields of any spin in $D > 4$. Our boundary terms contain a local combination of the bulk fields not involving any derivative along the direction normal to the regulating surface, thus guaranteeing that they do not spoil the variational principle. They however leave divergences localized at the boundary of null infinity that one can handle either by further tuning the boundary conditions or by adding appropriate ``corner'' counterterms. We also discuss how the latter approach relates to the renormalization of surface charges.

\section{Structure of the on-shell action} \label{sec:action}

To describe the dynamics of a free massless field of integer spin $s$, we consider a traceless tensor $\phi$ of rank $s$ and the Maxwell-like action of \cite{Skvortsov:2007kz, Campoleoni:2012th}. In Cartesian coordinates, it reads
\begin{equation} \label{Maxwell-like}
\begin{split}
	S = -\frac{1}{2} \int \text{d}^Dx \, \Big[
	\pr_\alpha \phi_{\mu_s}\, \pr^{\alpha} \phi^{\mu_s} 
	- s\, \lambda\, \pr_\alpha \phi_{\beta \mu_{s-1}} \pr^\beta \phi^{\alpha \mu_{s-1}} \\
    - s\,(1-\lambda)\, \pr\cdot \phi_{\mu_{s-1}}\, \pr\cdot \phi^{\mu_{s-1}} \Big] \, ,
\end{split}
\end{equation}
where $D$ is the dimension of spacetime, an index with a subscript denotes a set of symmetrized indices, \emph{e.g.}, $\phi_{\mu_s} \equiv \phi_{\mu_1 \cdots \mu_s}$, and we introduced the factor $\lambda \in \mathbb{R}$ to capture the possible rewritings of the action proportional to the boundary term 
\begin{equation} \label{cov_bnd_term}
    B_{\lambda} = 
    \frac{\lambda s}{2}\int \text{d}^Dx\, \partial_\alpha \left(\phi_{\beta\mu_{s-1}} \partial^\beta \phi^{\alpha \mu_{s-1}}-\partial^\beta\phi_{\beta\mu_{s-1}}  \phi^{\alpha \mu_{s-1}}\right) .
\end{equation}
For $s=1$ and $\lambda=1$, eq.~\eqref{Maxwell-like} gives the Maxwell action in its manifestly gauge-invariant form, while for $s=2$ one obtains the action of linearized unimodular gravity. The resulting on-shell action will anyway coincide with the one derived from the more customary Einstein or Fronsdal actions \cite{Fronsdal:1978rb} in the Bondi-like gauge \cite{Campoleoni:2017mbt, Campoleoni:2017qot, Campoleoni:2020ejn} considered below, in which the fields are traceless. 

The action \eqref{Maxwell-like} is invariant up to boundary terms under the gauge transformations\footnote{ Invariance of \eqref{Maxwell-like} actually holds even for fields and parameters subject to weaker trace conditions, including fully traceful ones. In those cases the spectrum becomes reducible and the equations of motion propagate additional particles of lower spins \cite{Campoleoni:2012th, Francia:2016weg}.}
\begin{equation} \label{gauge-symm}
\delta_\epsilon \phi_{\mu_s} = \pr_\mu \epsilon_{\mu_{s-1}} 
\quad \textrm{with} \quad
\pr\cdot\epsilon_{\mu_{s-2}} =0\,,\quad \epsilon_{\mu_{s-3}\alpha}{}^\alpha = 0 \, ,
\end{equation}
where repeated covariant or contravariant indices denote a symmetrization involving the minimum number of terms needed and without normalization factors. The equations of motion read 
\begin{equation} \label{eom_irr}
M_{\mu_s} := \Box \phi_{\mu_s} - \pr_{\mu} \pr\cdot \phi_{\mu_{s-1}} + \frac{2}{D+2s-4}\, \eta_{\mu\mu} \pr\cdot\pr\cdot \phi_{\mu_{s-2}} 
\approx
0 \,,
\end{equation}
where $\approx$ denotes equalities that hold on shell. 
Upon analyzing eq.~\eqref{eom_irr}, one finds that the gauge-invariant double divergence of~$\phi$ does not contain propagating degrees of freedom and on-shell satisfies
\begin{equation} \label{doublediv}
\pr \cdot \pr \cdot \phi_{\mu_{s-2}} \approx 0 \, ,
\end{equation}
possibly up to terms that do not vanish at the spacetime boundary and that we do not include in our solution space \cite{Skvortsov:2007kz, Campoleoni:2012th}.

The on-shell action takes the form
\begin{equation}\label{on-shell-action}
	S = \int \text{d}^Dx\, \mathcal{L}[\phi] \approx \frac{1}{2} \int \text{d}^Dx\,\partial_\alpha \hat{\theta}^\alpha[\phi,\phi]\,,
\end{equation}
where\footnote{For simplicity, in this section we work in Cartesian coordinates. In generic coordinates, both $\cL[\phi]$ and $\hat{\theta}^\alpha$ in \eqref{on-shell-action} are densities, thus including a factor $\sqrt{-g}$.}
\begin{equation} \label{bilinear0}
    \begin{split}
	\hat\theta^\alpha[\psi,\phi] &= -\, \phi_{\mu_s}\, \partial^{\alpha} \psi^{\mu_s} + s\, \lambda\, \phi_{\beta \mu_{s-1}} \partial^\beta \psi^{\alpha \mu_{s-1}}\\
	&\quad + s\,(1-\lambda)\, \phi^{\alpha\mu_{s-1}}\, \pr\cdot \psi_{\mu_{s-1}} \, ,
\end{split}
\end{equation}
which can also be cast in the form
\begin{equation} \label{bilinear}
\begin{split}
\hat{\theta}^\alpha[\psi,\phi] 
&= 
-\, \phi^{\mu_s} \Gamma^\alpha{}_{\mu_s} + s\, (1-\lambda) \Big[ \phi^\alpha{}_{\mu_{s-1}} \partial\cdot \psi^{\mu_{s-1}}
\\
&\quad - \phi^{\beta\mu_{s-1}} \partial_\beta \psi^\alpha{}_{\mu_{s-1}} \Big] \, ,
\end{split}
\end{equation}
with $\Gamma^\alpha{}_{\mu_s} = \pr^\alpha \psi_{\mu_s} - \pr_\mu \psi^\alpha{}_{\mu_{s-1}}$. This is the first de Wit--Freedman connection \cite{deWit:1979sib}, which for \mbox{$s=1$} gives the field strength and for~$s=2$ is proportional to the linearized Christoffel symbols.

Since we are considering a quadratic action, both the (Lee--Wald) presymplectic potential \cite{Lee:1990nz}, defined by
\begin{equation} \label{eq:deltaL}
    \delta \mathcal{L}[\phi] \approx \partial_\alpha \theta^\alpha[\phi,\delta\phi] \, ,
\end{equation}
and the boundary term that one obtains when considering a gauge variation of the Lagrangian, 
\begin{equation} \label{eq:gaugevar}
	\delta_\epsilon \cL[\phi] = \partial_\alpha B_\epsilon^\alpha[\phi] \, ,
\end{equation}
can be expressed in terms of \eqref{bilinear0}:
\begin{equation}\label{eq:defsthetaB}
    \theta^\alpha[\phi, \delta \phi] = \hat{\theta}^\alpha[\phi, \delta \phi] \, , \quad B_\epsilon^\alpha[\phi] = \hat{\theta}^\alpha[\delta_\epsilon \phi, \phi] \, .
\end{equation}
These two objects enter the Noether current
\begin{equation}\label{eq:Noether1}
    j_\epsilon^\alpha[\phi] = \theta^\alpha[\phi,\delta_\epsilon \phi] - B_\epsilon^\alpha[\phi]\,,
\end{equation}
which obeys Noether's second theorem
\begin{equation}\label{eq:Noether2}
    j_\epsilon^\alpha[\phi] \approx \partial_\beta \kappa_\epsilon^{\alpha\beta}[\phi]\,,
\end{equation}
where (see \ref{app:NoetherI}) 
\begin{equation}\label{eq:kappaalphabeta}
    \begin{split}
    \kappa_\epsilon^{\alpha\beta} &= s(s-1)\, {\epsilon_{\mu_{s-2}}}^{[\alpha} \partial \cdot \phi^{\beta]\mu_{s-2}} - s \, \epsilon_{\mu_{s-1}} \partial^{[\alpha} \phi^{\beta]\mu_{s-1}}\\
    &\quad + s \, \lambda \, \phi^{\mu_{s-1}[\alpha} \partial_\mu {\epsilon^{\beta]}}_{\mu_{s-2}} + s (1-\lambda)\,\partial^{[\alpha} \epsilon_{\mu_{s-1}} \phi^{\beta]\mu_{s-1}}
    \end{split}
\end{equation}
is the Noether two-form, and where square brackets denote antisymmetrization over the enclosed indices, with no overall normalization factor. This can also be cast in the equivalent form
\begin{equation}\label{eq:isolatinglambda}
    \kappa_\epsilon^{\alpha\beta} = \kappa_\epsilon^{\alpha\beta}\bigg|_{\lambda=0}
    + \lambda s\, \phi_{\mu_{s-1}}{}^{[\alpha} \delta_\epsilon\phi^{\beta]\mu_{s-1}}\,.
\end{equation}

Therefore, eq.~\eqref{eq:Noether2} implies that, on shell, the quantity \eqref{bilinear0} is symmetric in its arguments up to the divergence of an antisymmetric tensor (which we shall refer to as a corner term) when $\psi = \delta_\epsilon \phi$:
\begin{equation} \label{symmetry}
\hat{\theta}^\alpha[\phi, \delta_\epsilon \phi] \approx \hat{\theta}^\alpha[ \delta_\epsilon \phi , \phi ] + \partial_\beta \kappa_\epsilon^{\alpha\beta}[\phi] \, .
\end{equation}
This property will be important in the renormalization of the on-shell action discussed in sect.~\ref{sec:renormalization}.

\section{The Bondi-like gauge} \label{sec:Bondi}

To analyze the structure of the divergent terms in the on-shell action, we parameterize the Minkowski background using retarded Bondi coordinates as 
\begin{equation} \label{Bondi-coords}
    \text{d}s^2 =  - \text{d}u^2 - 2\, \text{d}u \, \text{d}r + r^2 \gamma_{ij} \text{d}x^i \text{d}x^j \, ,
\end{equation}
where $u = t -r$ and $\gamma_{ij}$ denotes the round metric on the codimension-2 unit celestial sphere, and where we impose the Bondi-like conditions \cite{Campoleoni:2017mbt, Campoleoni:2017qot, Campoleoni:2020ejn}
\begin{equation} \label{Bondi-like}
    \phi_{r\mu_{s-1}} = 0 \, , \quad \gamma^{ij} \phi_{ij\mu_{s-2}} = 0 \quad \Rightarrow \quad g^{\alpha\beta} \phi_{\alpha\beta\mu_{s-2}} = 0 \, .
\end{equation}
For simplicity, we thus fix both the coordinates of the background and the gauge, while commenting on this choice in the Conclusions. Notice that the conditions \eqref{Bondi-like} imply that the Fronsdal action reduces to the Maxwell-like action \eqref{Maxwell-like}. 

Following \cite{Campoleoni:2020ejn}, for any value of $D$ we further impose the boundary conditions
\begin{equation} \label{falloffs}
    \phi_{u_{s-k}i_k} := \phi_{\underbrace{\scriptstyle{u\,\cdots\,u}}_{s-k} i_1 \cdots i_k} = \cO(r^{k-1})
\end{equation}
on the remaining non-vanishing components. We also assume that the fields can be expanded in integer powers of the radial coordinate, without any $\log r$ contributions. When $D > 4$, the falloffs \eqref{falloffs} are overleading with respect to those typical of radiation, for which $\phi_{u_{s-k}i_k} = \cO(r^{-\frac{D-2}{2}+k})$. They however allow for residual gauge transformations generated by an arbitrary function on the celestial sphere, $T(\hat{x})$, which, following \cite{Campoleoni:2017mbt,  Campoleoni:2020ejn}, we refer to as higher-spin supertranslations.\footnote{Weaker boundary conditions leading to higher-spin generalizations of $\text{\textit{Diff}}(S^{D-2})$ superrotations have also been considered in \cite{Campoleoni:2020ejn}.} Actually, above the radiation order the equations of motion \eqref{eom_irr} only allow for pure-gauge configurations, so that the field components are bound to take the form \cite{Campoleoni:2020ejn}
\begin{equation} \label{on-shell}
    \phi_{u_{s-k}i_k} = r^{k-1} \frac{k(D+k-5)!}{s(D+s-5)!}\, (\cD\cdot)^{s-k} \hat{C}_{i_k}(\hat{x}) + \cO\left(r^{-\frac{D-2}{2}+k}\right) ,
\end{equation}
where the omitted terms do not depend on $T(\hat{x})$, and where
\begin{equation}\label{eq:hatCi1is}
    \hat{C}_{i_k}(\hat{x}) = \frac{1}{[(k-1)!]^2}\, \mathcal{D}_{\langle i_1}\cdots \mathcal{D}_{i_k\rangle} T(\hat{x})\,.
\end{equation}
In the latter expression, angular brackets denote the symmetric and traceless projection, while $\cD_i$ is the Levi--Civita covariant derivative on the celestial sphere. 

When $D>4$, in this gauge the space of solutions thus decomposes as
\begin{equation} \label{decomposition}
	\phi = \delta_\epsilon \phi + \phi_\text{rad} \, ,
\end{equation}
where $\delta_\epsilon \phi$ is a pure-gauge piece and $\phi_\text{rad}$ is a gauge-invariant one, encoding  information, \emph{e.g.}, about outgoing radiation and (subleading) Coulombic contributions. As highlighted by eq.~\eqref{on-shell}, the two terms in \eqref{decomposition} are neatly separated for $D>4$ because they appear at different orders in the large-$r$ expansion, with $\delta_\epsilon \phi$ being leading as $r\to\infty$, while for $D=4$ they mix.

\section{Renormalization of the action in Bondi-like gauge} \label{sec:renormalization}

The on-shell action obtained by evaluating \eqref{Maxwell-like} on the solution space \eqref{on-shell} diverges at null infinity. Our goal will be to show that the divergences can be canceled by adding a boundary term not affecting the variational principle. To identify it, we first define the regularized action by integrating the Lagrangian on a portion of spacetime up to a time-like surface at $r = R$:
\begin{equation}
S_{\text{reg}} = \int_{r \leq R} \text{d}^Dx \, \cL[\phi] \, .
\end{equation}
Using the property \eqref{symmetry} and the decomposition \eqref{decomposition}, the latter can be rewritten as
\begin{equation} \label{eq:Sreg}
\begin{split}
    S_\text{reg} &\approx \frac{1}{2} \int_{r = R} \text{d}^{D-1}x\, \Big(\, \hat{\theta}^r[\delta_\epsilon \phi,\delta_\epsilon \phi] + 2\,\hat{\theta}^r[\delta_\epsilon \phi,\phi_\text{rad}]\\
    &\quad + \hat{\theta}^r[\phi_\text{rad},\phi_\text{rad}] + \partial_\mu \kappa_\epsilon^{r\mu}[\phi]\, \Big) \, ,
\end{split}
\end{equation}
with 
\begin{equation} \label{theta^r}
\begin{split}
    & \hat{\theta}^r[\psi,\phi] = r^{D-2s-3} \sqrt{\gamma}\, \Big\{ - \phi^{i_s} \Big[ r\pr_r - (1-\lambda) s - r \pr_u \Big] \, \psi_{i_s} \\
    & \quad - s\,r\, \Big[ \lambda\, \phi^{i_{s-1}j} \cD_j \psi_{u i_{s-1}} + (1-\lambda)\, \phi_u{}^{i_{s-1}} \cD\cdot \psi_{i_{s-1}} \Big] \\ 
    & \quad + s\,r^2 \phi_u{}^{i_{s-1}} \Big[ r\pr_r - 2\lambda (s-1) + (1-\lambda) (D-2) \Big] \, \psi_{u i_{s-1}} \Big\} \, ,
\end{split}
\end{equation}
and where angular indices are raised and lowered using the metric $\gamma_{ij}$ on the celestial sphere.

We now focus on the limit of eq.~\eqref{theta^r} as $r = R \to \infty$ at fixed $u$, and on the corresponding patch in \eqref{eq:Sreg}. In this limit, one approaches the region of the boundary corresponding to future null infinity $\mathscr{I}^+$,  and our goal will be to identify a boundary term defined on the regulating surface canceling the divergences in $R$ of eq.~\eqref{eq:Sreg}. Notice that in this limit the last term in eq.~\eqref{eq:Sreg} gives a corner contribution localized at the future/past boundaries $\mathscr{I}^+_\pm$ of $\mathscr{I}^+$. A similar treatment apply to $\mathscr{I}^-$, while to discuss the renormalization of the action in the remaining regions of the boundary of Minkowski space it would be more efficient to consider other coordinate systems and we do not address this issue here. Eq.~\eqref{on-shell} then shows that $\hat{\theta}^r[\phi_\text{rad},\phi_\text{rad}]$ gives a contribution which remains finite in the limit $R \to \infty$ (at fixed~$u$) for any value of $D$. The terms in the first line of \eqref{eq:Sreg} and the corner term, depending on~$\kappa_\epsilon^{r\mu}[\phi]$, diverge instead in this limit when $D > 4$. 

Before discussing how to cancel these divergences, let us stress that for $s=1$ and $\lambda = 1$ one obtains
\begin{equation} \label{thetars1}
    \hat{\theta}^r[B, A] = - A_\mu (\partial^r B^\mu - \partial^\mu B^r) \,,
\end{equation}
involving in particular the field strength of $B^\mu$. This term vanishes when $B^\mu$ is pure gauge, thus \eqref{eq:Sreg} implies that the on-shell Maxwell action in flat space is manifestly finite up to corner terms. For $\lambda \neq 1$, one can regularize the action (again up to corner terms) by adding the covariant boundary term \eqref{cov_bnd_term} that gives back the manifestly gauge-invariant, $\lambda = 1$ form of the action. For $s>1$, the first \mbox{de Wit}--Freedman connection entering \eqref{bilinear} transforms instead as $\delta \Gamma^\alpha{}_{\mu_{s}} = - 2\, \pr^2{}_{\!\!\!\mu} \epsilon^\alpha{}_{\mu_{s-2}}$, and one has to add a boundary counterterm for any value of $\lambda$.

In view of \eqref{eq:Sreg}, we now specialize eq.~\eqref{theta^r} to field configurations in which $\psi = \delta_\epsilon \phi$ is a pure-gauge contribution involving only higher-spin supertranslations, as in eqs.~\eqref{on-shell} and \eqref{eq:hatCi1is}. Since the variation of each field component under supertranslations is homogeneous in $r$, in \eqref{theta^r} the operator $r\pr_r$ just produce a factor, \emph{i.e.}, $r\pr_r \delta_\epsilon \phi_{u_{s-k}i_k} = (k-1)\, \delta_\epsilon \phi_{u_{s-k}i_k}$. Moreover, the gauge invariance of the component $M_{ri_{s-1}}$ of the equations of motion (see eq.~(140) of \cite{Campoleoni:2017qot} for an explicit expression) implies
\begin{equation} \label{T_rel}
    (s-1) \, \mathcal{D} \cdot \delta_\epsilon \phi_{i_{s-1}} =rs(D+s-5)\,\delta_{\epsilon} \phi_{ui_{s-1}} \,.
\end{equation}
For $s=1$, eq.~\eqref{T_rel} gives $\delta_\epsilon A_u =0$. One then has (for $s\ge2$)
\begin{equation}\label{eq:thetarewritingepsilon}
\begin{split}
    & r^{2s+3-D}\,\hat{\theta}^r[\delta_\epsilon\phi, \phi] = \\ 
    & \ \sqrt{\gamma}\, \Big\{ (1-\lambda s)\, \phi^{i_s} \delta_\epsilon \phi_{i_s} + \frac{\lambda (s-1)}{D+s-5}\, \cD\cdot \phi^{i_{s-1}} \cD\cdot \delta_\epsilon \phi_{i_{s-1}} \\
    & \ \phantom{\sqrt{\gamma}} - \frac{s\, r^2}{s-1} \left[ \lambda s (s-1) + (1-\lambda) (D-4) \right] \phi_u{}^{i_{s-1}} \delta_\epsilon \phi_{ui_{s-1}} \Big\} \,,
\end{split}
\end{equation}
where we also integrated by parts a $\cD_i$, an operation which does not spoil the structure \eqref{eq:Sreg} of the on-shell action (we neglect total divergences on the sphere, whose integral vanishes).

Both contributions in the first line of eq.~\eqref{eq:Sreg} have this form and diverge as $\cO(R^{D-5})$. These divergences can however be canceled by defining the subtracted action
\begin{equation} \label{eq:Ssub}
	S_\text{sub} = S_\text{reg} + \int_{r=R} \text{d}^{D-1}x\,\mathcal{L}_\text{ct}\,,
\end{equation}
with a counterterm given by, for $s\ge2$,
\begin{align}\label{eq:Lct}
    \cL_\text{ct} & = \frac{\sqrt{-g}}{2R}\, \Big\{ (\lambda s - 1)\, \phi^{i_s}  \phi_{i_s} - \frac{R^2\lambda (s-1)}{D+s-5}\, \cD\cdot \phi^{i_{s-1}} \cD\cdot  \phi_{i_{s-1}} \nn \\
    &\quad + \frac{s}{s-1} \left[ \lambda s (s-1) + (1-\lambda) (D-4) \right] \phi_u{}^{i_{s-1}} \phi_{ui_{s-1}} \Big\} \,, 
\end{align}
where here we raised angular indices with the full metric $g^{ij}$ to absorb the corresponding powers of $r$. 
Instead, for $s=1$,
\begin{align}\label{eq:Lcts=1}
    \cL_\text{ct} & = \frac{\sqrt{-g}}{2R}\,(\lambda - 1) \left( A^{i}  A_{i} - 2R\,A_u\,\mathcal{D}\cdot A  \right) .
\end{align}
One can compare with the boundary term \eqref{cov_bnd_term}, which also renormalizes the action by making it manifestly gauge invariant, 
\begin{equation}\label{eq:Lcts=1cov}
    B_\lambda = -\frac{\lambda}{2R} \int_{r=R} \sqrt{-g}
    \left( A^{i}  A_{i} - 2R\,A_u\,\mathcal{D}\cdot A 
    + (D-2)\, A_u^2\right) .
\end{equation}
Eqs.~\eqref{eq:Lcts=1} and \eqref{eq:Lcts=1cov} are compatible since the last term in \eqref{eq:Lcts=1cov}, being gauge invariant, goes to zero as $R\to\infty$.

Notice that the counterterm \eqref{eq:Lct} is written in terms of bulk fields and it does not contain any derivative in $r$ of the fields, \emph{i.e.}, it does not involve derivatives in the direction normal to the regulating surface at $r = R$. As such, it is a boundary term that does not affect the variational principle. We were able to achieve this goal thanks to the option of letting all derivatives act on $\delta_\epsilon \phi$, which is guaranteed by the property \eqref{symmetry} of the on-shell action, and by the homogeneity in~$r$ of the variations of each field component under higher-spin supertranslations. Another crucial property of the counterterm~\eqref{eq:Lct} is that it only involves squares of each tensorial structure, which allows one to cancel both divergences in the first line of eq.~\eqref{eq:Sreg}. For~$D=4$, consistently with the finiteness of the on-shell action~\eqref{on-shell-action}, the counterterm vanishes in the limit $R \to \infty$. 

The subtracted action takes the form 
\begin{equation}\label{eq:sub}
\begin{split}
	S_\text{sub} 
    &\approx \frac{1}{2} \int_{r=R} \text{d}^{D-1}x \sqrt{\gamma}\, C^{i_s} \partial_u C_{i_s}
    \\ 
    &\quad - \frac{1}{2} \oint \text{d}^{D-2}x \, \kappa_\epsilon^{ur}[\phi] \bigg{|}_{u=-\infty}^{u=+\infty}+ \cO(R^{-1}) \,,
\end{split}
\end{equation}
where $C_{i_s}$ is the generalization of the $s=2$ shear tensor parameterizing the asymptotic solution space:
\begin{equation}
\phi_{i_s} = r^{s-1}\, \hat{C}_{i_s}(\hat{x}) + r^{-\frac{D+2s-2}{2}} C_{i_s}(u,\hat{x}) + \cO(r^{-\frac{D+2s-4}{2}}) \,.
\end{equation}
The tensor $C_{i_s}(u,\hat{x})$ is traceless but otherwise unconstrained, so that the first line of eq.~\eqref{eq:sub} is the generalization of the Ashtekar-Streubel structure \cite{Ashtekar:1981bq} to any massless bosonic field. It comes from the original on-shell action, since the counterterm vanishes as $\cO(R^{-1})$ when evaluated on radiation falloffs.

The second line of \eqref{eq:sub}, instead, still contains a potentially diverging corner contribution. The corner term coincides however with that entering Noether's second theorem, see eq.~\eqref{eq:kappaalphabeta}. The corresponding surface charge had already been discussed in ref.~\cite{Campoleoni:2020ejn}, focusing on $\lambda=0$.\footnote{See eq.~(A1) of \cite{Campoleoni:2017qot} and Footnote~5 of \cite{Campoleoni:2020ejn}. For comparison, let us recall that refs.~\cite{Campoleoni:2017qot,Campoleoni:2020ejn} used different normalization conventions for the two form:
$
        \kappa_\epsilon^{\alpha\beta} = s\, (\kappa_\epsilon^{\alpha\beta})_{\text{ref.~\cite{Campoleoni:2020ejn}}} = -s!\, (\kappa_\epsilon^{\alpha\beta})_{\text{ref.~\cite{Campoleoni:2017qot}}}
$.
Consistently, compared to (2.14) of \cite{Campoleoni:2020ejn}, the charge computed in \eqref{eq:CHARGE} below has an extra factor of $s$.
\label{footnote:normalization}
} 
However, as displayed by eq.~\eqref{eq:isolatinglambda}, the $\lambda$-dependent terms are proportional to $\phi^{\mu_{s-1}[\alpha} \delta_\epsilon \phi^{\beta]}_{\mu_{s-1}}$ and this vanishes in Bondi-like gauge for $\alpha=u$ and $\beta=r$, because $\phi^{u\mu_{s-1}}=-\phi_{r}{}^{\mu_{s-1}}=0$ and similarly for $\delta_\epsilon \phi^{u\mu_{s-1}}$. So, there is no new contribution to $\kappa_\epsilon^{ur}$ compared to \cite{Campoleoni:2020ejn}. As such, the corner contributions in the second line of \eqref{eq:sub} have the same structure as the supertranslation charge discussed in \cite{Campoleoni:2020ejn}. In that reference, it was shown that, for spin-$s$ supertranslations,
\begin{equation}\label{eq:magic}
    \oint \text{d}^{D-2}x \, \kappa^{ru}_{\epsilon_1}[\delta_{\epsilon_2} \phi] = 0 \, ,
\end{equation}
so that the (divergent) contributions that are ``quadratic'' in the pure-supertranslation part vanish identically. In fact, these contributions would also cancel in \eqref{eq:sub} because they are $u$-independent. Moreover, the surface charge had been shown to be finite when evaluated in a neighborhood of $\mathscr{I}^+_\pm$, say for $u < u_1$ and $u > u_2$ (with $u_1 < u_2$), under the assumption that in those regions there is no radiation and the fields attain a stationary configuration. A characterization of stationary solutions for fields of any spin is discussed in Appendix~D of \cite{Campoleoni:2020ejn}, and it amounts to consider configurations that near $\mathscr{I}^+_\pm$ behave like
\begin{equation}
\phi_{u_{s-k}i_k} = r^{3-D+k} \cU_{i_k}{}^{\!(k)}(u,\hat{x}) + \cO(r^{2-D+k}) 
\end{equation}
possibly up to a pure-gauge contribution of the same form as that displayed in eq.~\eqref{on-shell}.

Imposing these boundary conditions at $u \to \pm \infty$, the resulting finite renormalized action is 
\begin{equation}\label{eq:ren}
\begin{split}
 	S_\text{ren} &= \lim_{\substack{R \to \infty}} S_\text{sub} = \frac{1}{2} \int_{\mathscr{I}^+} \text{d}^{D-1}x\,\sqrt{\gamma}\,C^{i_s} \partial_u C_{i_s}
    \\
 	&+\frac{1}{2}\, s (-1)^{s}(D+s-4) \oint \text{d}^{D-2}x\,\sqrt{\gamma}\,T\,\mathcal{U}^{(0)}\bigg{|}_{u=-\infty}^{u=+\infty}\,.
\end{split}
 \end{equation}
Alternatively, one can avoid imposing stationarity of the fields at $u \to \pm \infty$ and add corner counterterms canceling the divergent contributions in the presymplectic potential. We shall discuss this approach in the next section. 

Let us now consider the variation of the subtracted action~\eqref{eq:sub} to compute the subtracted presymplectic potential, 
\begin{equation}
     \delta S_\mathrm{sub} \approx \int_{r = R} \text{d}^{D-1}x \, \theta^r_\mathrm{sub}[\phi,\delta \phi]
\end{equation}
where
\begin{equation}\label{eq:thetasubthetarLct}
    \theta^r_\mathrm{sub}[\phi,\delta \phi] = \theta^r[\phi,\delta \phi] + \delta \mathcal{L}_\mathrm{ct} \,.
\end{equation}
Evaluating this explicitly (see~\ref{app:thetaev}), we obtain
\begin{equation}\label{eq:thetasubev}
    \theta^r_\mathrm{sub}[\phi,\delta \phi] =  \theta^r[\phi_\mathrm{rad},\delta\phi_\mathrm{rad}] + \partial_u \kappa^{ru}_{\delta \epsilon}[\phi] \,.
\end{equation}
We can then define the renormalized presymplectic potential by taking the $R\to\infty$ limit in a non-radiative region, e.g. for $u<u_1$,
\begin{equation}
    \theta^r_\mathrm{ren}[\phi,\delta \phi] = \lim_{\substack{R \to \infty \\ u = \text{const}}} \theta^r_\mathrm{sub}[\phi,\delta \phi]\,,
\end{equation}
for which we find
\begin{equation}
    \begin{split}
    &\theta^r_\mathrm{ren}[\phi,\delta \phi] 
    \\
    &\approx \sqrt{\gamma}\left( \delta C^{i_s}\,\partial_u C_{i_s} + s(-1)^{s}(D+s-4)\, \partial_u \left( \delta T \, \mathcal{U}^{(0)} \right) \right) .
    \end{split}
\end{equation}
The latter can be used to obtain the renormalized supertranslation charge. Taking into account that $\delta_\epsilon C_{i_s}=0$, $\delta_\epsilon \mathcal{U}^{(0)}=0$ and $\delta_\epsilon T = T$ under spin-$s$ supertranslations,
\begin{equation}\label{eq:THETARENr}
    \theta^r_\text{ren}[\phi,\delta_\epsilon \phi] =  s(-1)^{s}(D+s-4)\,
    \sqrt{\gamma}\, T \, \partial_u\mathcal{U}^{(0)} \, .
\end{equation}
Since the charge is $\delta$-integrable in our linearized setup, integrating \eqref{eq:THETARENr} over a hypersurface with boundary at $\mathscr{I}^{+}_-$, yields
\begin{equation}\label{eq:CHARGE}
    \mathcal{Q}_T = \oint \text{d}^{D-2}x\,s(-1)^{s-1}(D+s-4)\, \sqrt{\gamma}\, T \, \mathcal{U}^{(0)}\, ,
\end{equation}
in agreement with \cite{Campoleoni:2020ejn} (see footnote~\ref{footnote:normalization}).

\section{Corner terms}\label{sec:corner}

In the previous section we discussed how, starting from the formal action \eqref{Maxwell-like}, one can obtain a finite action at null infinity by adding the boundary counterterm \eqref{eq:Lct} and imposing stationary field configurations at corners of $\mathscr{I}^+$, that is, for $u \to \pm \infty$. In this section, we follow an alternative route by identifying corner counterterms. 
To this end, it is natural to investigate the symplectic structure associated with eq.~\eqref{eq:deltaL}, because the technical steps are the same as those required to renormalize surface charges at finite $u$.

We start by recalling that the (Lee--Wald) presymplectic potential $\theta^{\alpha}[\phi, \delta \phi]$ admits two types of ambiguities \cite{Wald:1993nt},
\begin{equation} \label{eq:IWambi}
    \theta^\alpha[\phi,\delta \phi] \to \theta^\alpha[\phi,\delta \phi] + \delta Z^\alpha[\phi] + \partial_\beta Y^{\alpha\beta}[\phi,\delta \phi] \, ,
\end{equation}
where $Y^{\alpha\beta}$ is an antisymmetric term linear in the field variations. The first ambiguity in \eqref{eq:IWambi} corresponds to adding a boundary term to the Lagrangian, $\mathcal{L} \to \mathcal{L} + \partial_\alpha Z^\alpha$ , and does not contribute to the associated presymplectic form,
\begin{equation}
    \omega^\alpha_\epsilon = \delta \theta^\alpha[\phi,\delta_\epsilon \phi] - \delta_\epsilon \theta^\alpha[\phi,\delta \phi] \, ,
\end{equation}
which in turn determines the Iyer--Wald surface charge density:
\begin{equation}\label{eq:identitytheta}
	  \omega^\alpha_\epsilon \approx \partial_\beta\kappa^{\alpha\beta}_\epsilon[\delta \phi]\,.
\end{equation}
Notice that for the quadratic theory \eqref{Maxwell-like}, in which the gauge parameters are field independent, the last relation \eqref{eq:identitytheta} can be $\delta$-integrated to eqs.~\eqref{eq:Noether2} and \eqref{eq:kappaalphabeta}. We also note that the boundary modification~$Z^r$ may correspond to the counterterm \eqref{eq:Lct}. The second ambiguity in \eqref{eq:IWambi} arises from the fact that $\theta^\alpha$ appears as a boundary term in $\delta \mathcal{L}$. This term, called corner term, affects~$\omega^\alpha_\epsilon$,
\begin{equation}
    \omega^\alpha_\epsilon \to \omega^\alpha_\epsilon + \partial_\beta \Big( \delta Y^{\alpha\beta}[\phi,\delta_\epsilon \phi] - \delta_\epsilon Y^{\alpha\beta}[\phi,\delta \phi] \Big) \, ,
\end{equation}
and thus also modifies the charge two-form. As far as the renormalization of the charge is concerned, only the corner term plays a role. For this reason, we now focus on the latter and will relate it to the corner integral appearing in eq.~\eqref{eq:Sreg}.

Actually, there is a simple prescription in the choice of $Y^{\alpha\beta}$ that cancels all charge divergences for generic values of $u$. 
Focusing on the component that eventually enters the expression for the charge, we can choose \cite{McNees:2023tus,McNees:2024iyu}
\begin{equation} \label{eq:s1Cur2}
    Y^{ru} = \int \mathrm{d}r \, \theta^u \, ,
\end{equation}
and thus define the subtraction for the radial component of the symplectic structure according to
\begin{equation}\label{eq:theta_sub}
    \theta^r_\text{sub} = \theta^r + \delta Z^r + \partial_u \int \mathrm{d}r \, \theta^u \, .
\end{equation}
To see why this leads to finite charges in the asymptotic limit~$r \to \infty$, let us note that, owing to \eqref{eq:deltaL},
\begin{equation}
    \partial_r \theta^r_\text{sub} = \partial_r \theta^r + \delta \left( \partial_r Z^r \right) + \partial_u \theta^u \approx \delta \left( \mathcal{L} + \partial_r Z^r \right) - \partial_i \theta^i.
\end{equation}
Thus we see that the $r$-dependent, and in particular $r$-divergent, terms of $\theta_\text{sub}^r$ reduce to $\delta$-exact contributions or total divergences on the sphere, which vanish when calculating the presymplectic form and hence the charge. In this way, the subtraction defined in \eqref{eq:theta_sub} ensures a finite $r\to\infty$ limit and thus provides a renormalized charge for generic spins and dimensions, regardless of~$Z^\alpha$.

For instance, for $s = 1$ and $\lambda=1$, we have 
\begin{equation}
    \theta^{\alpha}[A,\delta A] = -\sqrt{-g} F^{\alpha\nu}\delta A_\nu\,,\qquad \kappa_\epsilon^{\alpha \beta} = -\sqrt{-g} F^{\alpha\beta}\,\epsilon\,,
\end{equation}
and thus
\begin{equation}
    \theta^u[A,\delta A] = r^{D-4} \sqrt{\gamma} \, \partial_r A_i\, \gamma^{ij} \delta A_j \, .
\end{equation}
This leads to the following expression for the divergent terms,
\begin{equation} \label{eq:s1Cur3}
    Y^{ur} = \sum_{k = \frac{D-2}{2}}^{D-4} r^{D-3-k} \frac{k-1}{D-3-k} \sqrt{\gamma} \, A_i^{(k-1)} \gamma^{ij} \delta A^{(0)}_j 
    +
    \mathcal{O}(1)\, ,
\end{equation}
where we denoted by $A^{(k)}_i$ the coefficients in the radial expansion of the on-shell fields, \emph{i.e.}, $A_i = \sum_k r^{-k} A^{(k)}_i$. This choice ensures the cancellation of the divergent terms in the surface charge, which we can write as follows after integrating by parts and using the equations of motion,
\begin{equation}
\begin{split}
    &\oint \kappa_\epsilon^{ur} \text{d}^{D-2}x
	=r^{D-2}  \oint  \text{d}^{D-2}x\,\sqrt{\gamma}\,
	F_{ur} T  
    \\
    =
	&-\sum_{k=\frac{D-2}{2}}^{D-4} r^{D-3-k}  \frac{k-1}{D-3-k} \oint \text{d}^{D-2}x\,\sqrt{\gamma}\,  A_i^{(k-1)} \gamma^{ij} \partial_j T
    \\
    &+(D-3)\oint \text{d}^{D-2}x \sqrt{\gamma}\,T\,A_u^{(D-3)}
    +\mathcal{O}(r^{-1})\,.
\end{split}
\end{equation}
We note that, in this case, choosing the finite part of the counterterm \eqref{eq:s1Cur3} to vanish, the resulting charge coincides with the result \eqref{eq:CHARGE} obtained in the previous section. Therefore, with this prescription, an appropriate choice of ambiguities that renormalizes the symplectic structure is given by the boundary modification \eqref{eq:thetasubthetarLct}, $Z^r = \mathcal{L}_\mathrm{ct}$, together with the corner term \eqref{eq:s1Cur2}.\footnote{Let us observe that applying the Comp\`ere--Marolf prescription \cite{Compere:2008us} starting from $Z^r = \mathcal{L}_\text{ct}$ would not yield any nontrivial corner counterterm for the presymplectic potential, because there are no $\partial_u$ in it.}

We note that, while it is quite general, the subtraction defined by \eqref{eq:theta_sub} is not manifestly local in terms of bulk fields. Indeed, the general form \eqref{eq:s1Cur2} involves an integral with respect to the radial coordinate, while the explicit form \eqref{eq:s1Cur3} involves direct dependence on the coefficients of the $1/r$ expansion, as opposed to the bulk fields.

Just as we have done for the boundary counterterm in the last section, one might thus ask whether it is still possible to express the corner counterterm \eqref{eq:s1Cur3} in terms of the bulk fields in a local way. An obstruction is represented by the fact that the most divergent term in \eqref{eq:s1Cur3} bears the form
\begin{equation}\label{eq:leading-singularity}
    Y^{ur} = r^{\frac{D-4}{2}} \sqrt{\gamma} \, A_{i}^{(\frac{D-4}{2})}\,\gamma^{ij}\,\delta A_j^{(0)} + \mathcal{O}(r^{\frac{D-6}{2}}) \, .
\end{equation}
The only two structures that can be built with the bulk field ${A}_\mu$ and do not involve angular or time derivatives, like the leading singularity displayed in \eqref{eq:leading-singularity}, are ${A}_i \gamma^{ij} \delta {A}_j$, ${A}_u \delta {A}_u$. However, these are total variations and hence do not contribute to the presymplectic form. On the other hand, one could construct one such counterterm by allowing for radial derivatives, but this is perturbatively equivalent to an expression such as \eqref{eq:s1Cur3} itself which involves explicit coefficients of the $1/r$ expansion.

Thus, our analysis indicates that the corner renormalization performed at the level of the symplectic potential for generic $u$ involves terms that are either nonlocal or involve explicit dependence on the $1/r$ expansion coefficients. Naturally, if we now once again assume that radiation vanishes as $u \to -\infty$, in particular before some reference $u_1$, all seemingly divergent terms in the presymplectic potential become harmless as they vanish when calculating the charge at $\mathscr{I}^+_-$ \cite{Campoleoni:2020ejn,Capone:2023roc}.

Notably, the corner counterterm \eqref{eq:s1Cur2} also leaves the finite term formally unspecified, and as such would require additional criteria to identify the finite remainder after the subtraction in a unique way. For instance, one can employ a ``minimal subtraction'' scheme, in which only divergent terms in $r$ are included in the counterterm as above, or impose additional properties on the allowed counterterms, see \cite{Odak:2022ndm, Capone:2023roc, Riello:2024uvs, Rignon-Bret:2024wlu}. We leave the investigation of this point to future work.

\section{Conclusions} \label{sec:conclusions}

We studied the renormalization of the divergent on-shell action and surface charges that one encounters when considering free massless fields of arbitrary spin with falloffs at null infinity allowing for supertranslation-like asymptotic symmetries. Our treatment applies to any spacetime dimension~$D > 4$. For each value of the spin $s$, we identified the boundary counterterm \eqref{eq:Lct}, which contains a local combination of the bulk fields not involving any derivative along the normal to the regulating surface. Imposing suitable boundary conditions in far past and future, this counterterm gives a finite action at null infinity without affecting the variational principle. We also discussed in sect.~\ref{sec:corner} how one could weaken these corner conditions, noticing however that this requires to introduce corner counterterms that are either non-local in the bulk fields or include derivatives in the direction normal to the regulating surface. 

The construction of our local boundary counterterm relies on the property \eqref{symmetry} of the linearized action, as well as on the homogeneity in the radial coordinate of (higher-spin) supertranslations, see eq.~\eqref{on-shell}. The latter property is not shared by superrotations and their higher-spin  counterparts \cite{Campiglia:2014yka, Capone:2019aiy, Colferai:2020rte, Capone:2021ouo, Campoleoni:2020ejn} and conceivably would not hold for higher-spin supertranslations too, whenever non-linearities are included. Building a local counterterm also for asymptotic solution spaces allowing for these symmetries or including $\log r$ contributions (see, \emph{e.g.}, \cite{Geiller:2024ryw, Fuentealba:2024lll, Fuentealba:2025ekj}) would thus require further work. Similarly, a proper generalization of the above properties could be instrumental to investigate the nonlinear Einstein theory. 

The existence of a boundary counterterm containing a local combination of the fields was not obvious a priori, especially in view of the difficulties to achieve the same for the corner terms that one needs to renormalize the charges. On the other hand, in our approach we fixed both the coordinates of the Minkowski background and the gauge. It would be interesting to reconsider our results so as to establish a renormalization procedure encompassing covariance with respect to both diffeomorphisms and linearized gauge transformations, possibly resorting to conformal compactification methods \cite{Geroch:1977big, Ashtekar:2014zsa} or to the cohomological setup of the variational bicomplex \cite{Compere:2018aar}. We defer such investigations to future work, but we expect the asymptotic solution space to exhibit properties analogous to \eqref{decomposition} when one imposes gauge-fixing conditions like those in eq.~\eqref{Bondi-like} also in other coordinates systems. 
Combining a diffeomorphism with a compensating gauge transformation, one should thus be able to follow similar steps as those discussed above to renormalize the action. 

Another useful intermediate step would be to restore covariance at least on each regulating surface, in the spirit of what has been done in gravity in \cite{Campoleoni:2022wmf, Campoleoni:2023fug, Fiorucci:2025twa, Hartong:2025jpp}. As discussed in \ref{app:covariance}, this could be achieved, \emph{e.g.}, by parameterizing the background in an ADM form, thus possibly disclosing the underlying Carrollian geometry of the boundary, as in the mentioned references. More in general, it will be interesting to identify a procedure to derive boundary correlators from our on-shell actions and non-linear generalizations thereof, in the spirit of holographic renormalization in AdS. 

\section*{Acknowledgments}

We would like to thank Federico Capone, Luca Ciambelli, Sucheta Majumdar, Chrysoula Markou and C\'eline Zwikel for useful discussions. A.C.\ is a research associate of the Fonds de la Recherche Scientifique -- FNRS. His work was supported by FNRS under the grant T.0047.24. A.D.\ was supported by a National Lottery Fellowship of the Belgian American Educational Foundation. We thank the Galileo Galilei Institute for Theoretical Physics for the hospitality and the INFN for partial support during the completion of this work. We also thank UMONS and Rome III U.\ for the hospitality extended to us, at various stages during the preparation of this work.

\appendix

\section{The Noether two-form} \label{app:NoetherI}

We would like to compute the Noether current corresponding to the invariance of the action \eqref{Maxwell-like} under the gauge transformation \eqref{gauge-symm}. Let us write the Lagrangian as
\begin{equation} \label{Lagrlambda}
{\cal L} = \lambda \, {\cal L}_1 + (1 - \lambda)\,  {\cal L}_2 \, ,
\end{equation}
where
\begin{align}
& {\cal L}_1 =  -\frac{1}{2}\, 
	\pr_\rho \phi_{\mu_s}\, \pr^{\rho} \phi^{\mu_s} 
	+ \frac{s}{2}\, \pr_\rho \phi_{\sigma \mu_{s-1}} \pr^\sigma \phi^{\rho \mu_{s-1}}\, , \\
& {\cal L}_2 = -\frac{1}{2}\, 
	\pr_\rho \phi_{\mu_s}\, \pr^{\rho} \phi^{\mu_s} 
	+ \frac{s}{2}\, \pr\cdot \phi_{\mu_{s-1}}\, \pr\cdot \phi^{\mu_{s-1}}\, .
\end{align}
We can compute the quantity $\hat{\theta}^{\alpha}[\psi,\phi]$ defined in \eqref{bilinear0} and the corresponding current separately for each term ${\cal L}_1$ and ${\cal L}_2$ and then take their linear combination to get a result valid for any value of $\lambda$.

For ${\cal L}_1$ one finds
\begin{equation}
    \hat{\theta}^{\alpha}_1 [\psi, \phi] = \phi_{\mu_s} \, \partial^{\mu} \psi^{\alpha \mu_{s-1}} \, - \, \phi_{\mu_s} \, \partial^{\alpha} \psi^{\mu_s} \, ,
\end{equation}
which, following \cite{Iyer:1994ys}, gives the Noether current 
\begin{equation}
\begin{split}
j^{\alpha}_1 & = \hat{\theta}^{\alpha}_1 [\phi, \delta_{\epsilon} \phi] - \hat{\theta}^{\alpha}_1 [\delta_{\epsilon} \phi, \phi] \\
& = \pr_{\mu} \epsilon_{\mu_{s-1}} \pr^{\mu} \phi^{\alpha \mu_{s-1}} - \pr_{\mu} \epsilon_{\mu_{s-1}} \pr^{\alpha} \phi^{\mu_s} - 2 \, \pr^2{}_{\!\!\!\mu}\,  \epsilon^{\alpha}{}_{\mu_{s-2}} \phi^{\mu_s}\, ,
\end{split}
\end{equation}
where $\pr^2_{\mu}$ denotes a double gradient involving the minimal number of terms required for full symmetrization, {\it e.g.}, the square gradient of a vector,  $\pr^2{}_{\!\!\!\mu} A_{\mu}$, stands for the combination  $\pr_{\alpha} \pr_{\beta} A_{\gamma} + \pr_{\gamma} \pr_{\alpha} A_{\beta} + \pr_{\beta} \pr_{\gamma} A_{\alpha}$. Manipulating the sum of the first two terms one obtains
\begin{equation}
 \begin{split}
& \pr_{\mu} \epsilon_{\mu_{s-1}} \pr^{\mu} \phi^{\alpha \mu_{s-1}} - \pr_{\mu} \epsilon_{\mu_{s-1}} \pr^{\alpha} \phi^{\mu_s} = s\pr_{\beta}\big[\epsilon_{\mu_{s-1}} \big( \pr^{\beta} \phi^{\alpha \mu_{s-1}} +  \pr^{\mu} \phi^{\alpha \beta \mu_{s-2}} \\
& - \pr^{\alpha} \phi^{\beta \mu_{s-1}}\big)\big] - s \epsilon_{\mu_{s-1}} \big(\Box \phi^{\alpha \mu_{s-1}} + \pr_{\mu} \pr \cdot \phi^{\alpha \mu_{s-2}} - \pr^{\alpha} \pr \cdot \phi^{\mu_{s-1}}\big) \, , 
\end{split}
\end{equation}
while the third term can be written as
\begin{equation}
\begin{split}
- 2 \, \pr^2_{\mu}\,  \epsilon^{\alpha}{}_{\mu_{s-2}} \phi^{\mu_s}\, & = -s (s-1) \big[\pr \cdot \pr \cdot \phi_{\mu_{s-2}} \epsilon^{\alpha \mu_{s-2}} \\
& + \pr_{\beta} \big(\phi^{\beta \gamma \mu_{s-2}}\pr_{\gamma} \epsilon ^{\alpha}{}_{\mu_{s-2}} - \pr \cdot \phi^{\beta \mu_{s-2}}\epsilon ^{\alpha}{}_{\mu_{s-2}} \big) \big]\, .
\end{split}
\end{equation}
Putting these together one finds $j^{\alpha}_1 \approx \pr_{\beta} \kappa^{\alpha \beta}_1$ with
\begin{equation}
\kappa^{\alpha \beta}_1 = s \phi^{\mu_{s-1} [\alpha }\pr_{\mu} \epsilon^{\beta]}{}_{\mu_{s-2}} + s (s-1) 
\epsilon_{\mu_{s-2}}{}^{[\alpha} \pr \cdot \phi^{\beta] \mu_{s-2}} - s \epsilon_{\mu_{s-1}} \pr^{[\alpha} \phi^{\beta] \mu_{s-1}}\, ,
\end{equation}
where we took into account that eqs.~\eqref{eom_irr} and \eqref{doublediv}, in combination with the tracelessness of the gauge parameter, imply
\begin{equation} \label{onshell}
 \begin{split}  
& \pr \cdot \pr \cdot \phi_{\mu_{s-2}} \approx 0\, ,\\
& \epsilon^{\mu_{s-1}} \big (\Box \phi_{\mu_s} - \pr_{\mu} \pr\cdot \phi_{\mu_{s-1}}\big) \approx 0 \, .
 \end{split}
\end{equation}

We can now perform the same computation for ${\cal L}_2$. 
One finds
\begin{equation}
\hat{\theta}^{\alpha}_2 [\psi, \phi] = s\, \phi^{\alpha \, \mu_{s-1}} \, \pr \cdot \psi_{\mu_{s-1}} \, - \, \phi^{\mu_s} \, \pr^{\alpha} \psi_{\mu_s} \, ,
\end{equation}
which gives the Noether current 
\begin{equation} \label{omega}
\begin{split}
 j^{\alpha}_2 \, = & \, s\big(\partial^{\alpha} \epsilon^{\mu_{s-1}} + \partial^{\mu} \epsilon^{\alpha \mu_{s-2}}\big)  \pr \cdot \phi_{\mu_{s-1}}\, - \, \partial_{\mu} \epsilon_{\mu_{s-1}} \partial^{\alpha}  \phi^{\mu_s}   \\
 & - s \phi^{\alpha \mu_{s-1}} \Box  \epsilon_{\mu_{s-1}} \, + \, \phi^{\mu_s} \pr^{\alpha} \pr_{\mu} \epsilon_{\mu_{s-1}}\, .
 \end{split}
\end{equation}
Again, the idea is to express the current as a total derivative up to terms that vanish on shell as in \eqref{onshell}. In particular,  integrating by parts twice the term involving the d'Alembertian one obtains
\begin{equation}
\phi^{\alpha \mu_{s-1}} \Box  \epsilon_{\mu_{s-1}} = \Box \phi^{\alpha \mu_{s-1}} \epsilon_{\mu_{s-1}} \, + \, \partial_\beta \left(\phi^{\alpha \mu_{s-1}} \partial^\beta \epsilon_{\mu_{s-1}} - \pr^\beta \phi^{\alpha \mu_{s-1}} \epsilon_{\mu_{s-1}} \right) 
\end{equation}
where $\Box \phi^{\alpha \mu_{s-1}}$ can be completed to reproduce the equations of motion.  After some additional manipulations on the other terms, one eventually gets 
\begin{align}
\nonumber 
j^{\alpha}_2 & =   
 \partial_\beta \kappa^{\alpha \beta}_2
+s \epsilon^{\mu_{s-1}} \big(\Box \phi_{\alpha \mu_{s-1}} - \pr_{\alpha} \pr\cdot \phi_{\mu_{s-q}} - \pr_{\mu} \pr\cdot \phi_{\alpha \mu_{s-2}}\big) \\
\nonumber
 &\quad + s (s-1) \epsilon^{\alpha \mu_{s-2}} \partial \cdot \partial \cdot \phi_{\mu_{s-2}} \approx \partial_\beta \kappa^{[\alpha \beta]}_2\, ,
 \label{eq:omegakappa}
\end{align}
where 
\begin{equation}
\kappa^{\alpha \beta}_2 = s(s-1) \epsilon_{\mu_{s-2}}{}^{[\alpha} \partial \cdot \phi^{\beta] \mu_{s-2} }
 \, + \, s \pr^{[\alpha} \epsilon_{\mu_{s-1}} \phi^{\beta] \mu_{s-1}} 
\, - \, 
s \epsilon_{\mu_{s-1}} \partial^{[\alpha} \phi^{\beta] \mu_{s-1}}  \, .
\end{equation}
Altogether, the Noether current for the Lagrangian \eqref{Lagrlambda} is 
$j^{\alpha}_\epsilon \approx \partial_\beta \kappa_\epsilon^{\alpha \beta}$,
with
\begin{equation}
    \begin{split}
    \kappa_\epsilon^{\alpha \beta} &= \lambda \kappa^{\alpha \beta}_1 + (1-\lambda) \kappa^{\alpha \beta}_2 \\&= s(s-1) {\epsilon_{\mu_{s-2}}}^{[\alpha} \partial \cdot \phi^{\beta]\mu_{s-2}} - s \, \epsilon_{\mu_{s-1}} \partial^{[\alpha} \phi^{\beta]\mu_{s-1}}\\
    &\quad + s \, \lambda \, \phi^{\mu_{s-1}[\alpha} \partial_\mu {\epsilon^{\beta]}}_{\mu_{s-2}} + s (1-\lambda) \partial^{[\alpha} \epsilon_{\mu_{s-1}} \phi^{\beta]\mu_{s-1}}\,.
    \end{split}
\end{equation}
Let us note that the $\lambda$-dependent part of $\kappa^{\alpha\beta}_\epsilon$ can be rewritten in a simple way,
\begin{equation}
    \kappa_\epsilon^{\alpha\beta} = \kappa_2^{\alpha\beta}
    + \lambda s\, \phi_{\mu_{s-1}}{}^{[\alpha} \delta_\epsilon\phi^{\beta]\mu_{s-1}}\,.
\end{equation}
This is consistent with the fact that the term proportional to $\lambda$ in the Lagrangian \eqref{Lagrlambda} is a total derivative
\begin{equation}
    \mathcal{L}_1-\mathcal{L}_2 = \partial_\alpha \mathcal{E}^\alpha
\end{equation}
with
\begin{equation}
    \mathcal{E}^\alpha[\phi] = \frac{s}{2}\left(\phi_{\beta\mu_{s-1}} \partial^\beta \phi^{\alpha \mu_{s-1}}-\partial^\beta\phi_{\beta\mu_{s-1}}  \phi^{\alpha \mu_{s-1}}\right)
\end{equation}
and
\begin{equation}
    \theta^{\alpha}_1[\phi,\delta\phi]
    -
    \theta^{\alpha}_2[\phi,\delta\phi]
    =
    \delta \mathcal{E}^\alpha[\phi]
    +
    \frac{s}{2}\,\partial_\beta\left(\phi_{\mu_{s-1}}{}^{[\alpha}\delta\phi^{\beta]\mu_{s-1}}\right).
\end{equation}

\section{Evaluating \texorpdfstring{$\theta_\text{sub}$}{theta-sub}}
\label{app:thetaev}

To evaluate \eqref{eq:thetasubthetarLct} explicitly, let us manipulate the two ingredients on the right-hand side with the goal to highlight the cancellation of the divergent terms. We begin by noting that the counterterm Lagrangian \eqref{eq:Lct} was constructed in such a way as to cancel the first two terms on the right-hand side of \eqref{eq:Sreg}, and as such it satisfies (up to total divergences on the sphere and up to terms that eventually vanish as $R\to\infty$)
\begin{equation}\label{eq:Lctback}
\begin{split}
    \mathcal{L}_\mathrm{ct} 
    &\approx 
    -\hat{\theta}^r[\delta_\epsilon \phi,\phi_\text{rad}] - \tfrac{1}{2} \hat{\theta}^r[\delta_\epsilon \phi,\delta_\epsilon \phi]
    \\
    &\approx
    - \hat{\theta}^r[\delta_\epsilon \phi,\phi] + \tfrac{1}{2} \hat{\theta}^r[\delta_\epsilon \phi,\delta_\epsilon \phi]\,,
\end{split}
\end{equation}
where we used the decomposition \eqref{decomposition} of the solution space. Similarly, the ``bare'' presymplectic potential defined by \eqref{eq:defsthetaB} can be manipulated as follows,
\begin{equation}
\begin{split}
\theta^{\alpha}[\phi,\delta \phi] 
&\approx
\hat \theta^{\alpha}[\delta_\epsilon \phi,\delta \phi] 
+
\hat\theta^{\alpha}[\phi_\text{rad},\delta_{\delta \epsilon} \phi] 
+
\hat\theta^{\alpha}[\phi_\text{rad},\delta \phi_\text{rad}] 
\\
&\approx 
\hat{\theta}^{\alpha}[\delta_\epsilon\phi,\delta\phi] + \hat{\theta}^{\alpha}[\phi,\delta_{\delta\epsilon}\phi] 
- \hat{\theta}^{\alpha}[\delta_{\epsilon}\phi,\delta_{\delta\epsilon}
\phi]
\\
&
\quad +
\hat\theta^{\alpha}[\phi_\text{rad},\delta \phi_\text{rad}]\,,
\end{split}
\end{equation}
where we used again \eqref{decomposition} as well as the fact that $\delta(\delta_\epsilon \phi_{\mu_s}) = \partial_\mu \delta\epsilon_{\mu_{s-1}}=\delta_{\delta\epsilon} \phi_{\mu_s}$. Applying the second Noether theorem in the form \eqref{symmetry}, we then have
\begin{equation}
    \begin{split}
    \theta^{\alpha}[\phi,\delta \phi]
    &\approx 
    \hat{\theta}^{\alpha}[\delta_\epsilon\phi,\delta\phi] + \hat{\theta}^{\alpha}[\delta_{\delta\epsilon}\phi,\phi]
    - \hat{\theta}^{\alpha}[\delta_{\epsilon}\phi,\delta_{\delta\epsilon}
\phi]
\\
&
\quad +
\hat\theta^{\alpha}[\phi_\text{rad},\delta \phi_\text{rad}]
+
\partial_\beta
\kappa^{\alpha\beta}_{\delta\epsilon}[\phi]\,.
    \end{split}
\end{equation}
Finally, we focus on the $\alpha=r$ component and note that, again by \eqref{symmetry}, 
\begin{equation}
\begin{split}
&
\hat{\theta}^r[\delta_\epsilon \phi, \delta_{\delta\epsilon}\phi]
    -
    \hat{\theta}^r[\delta_{\delta\epsilon} \phi, \delta_{\epsilon}\phi] \approx
\partial_\beta \kappa^{r\beta}_{\delta\epsilon}[\delta_\epsilon \phi]   
\approx 
\partial_i \kappa^{ri}_{\delta\epsilon}[\delta_\epsilon \phi] \,,
\end{split}
\end{equation}
where in the last equality we have employed \eqref{eq:magic}. This allows us to recast $\theta^r[\phi,\delta\phi]$ in the following way, up to total divergences on the sphere,
\begin{equation}\label{eq:thetaback}
\begin{split}
    \theta^r[\phi,\delta \phi] 
    &\approx \delta \left( \hat{\theta}^r[\delta_\epsilon \phi,\phi] - \frac{1}{2} \hat{\theta}^r[\delta_\epsilon \phi,\delta_\epsilon \phi] \right) 
    \\
    &
    \quad + \theta^r[\phi_\mathrm{rad},\delta\phi_\mathrm{rad}] + \partial_u \kappa^{ru}_{\delta \epsilon}[\phi]\,.
\end{split}
\end{equation}
By \eqref{eq:Lctback}, we see that the terms in the first line of \eqref{eq:thetaback}  cancel out in $\theta^{r}[\phi,\delta \phi]+ \delta \mathcal{L}_\text{ct}$, leading from \eqref{eq:thetasubthetarLct} to \eqref{eq:thetasubev} in the main body of the text.

\section{Covariance under boundary diffeomorphisms} \label{app:covariance}

Holographic renormalization would require, in particular, that a boundary counterterm be covariant with respect to reparametrizations of the regulating surface at $r = R$. Drawing inspiration from the boundary-covariant gauges \cite{Hartong:2015usd,Campoleoni:2022wmf, Campoleoni:2023fug, Hartong:2025jpp} used in nonlinear gravity, this property can be made manifest using an ADM-like foliation of Minkowski spacetime in terms of time-like surfaces,
\begin{equation} 
    \text{d}s^2 = \text{d}r^2 + h_{ab}\, \big( \text{d}x^a + N^a \text{d}r \big) \big( \text{d}x^b + N^b \text{d}r \big) \, ,    
\end{equation}
where $x^a$ are the coordinates and $h_{ab}$ is the induced metric on each time-like sheet.
Indeed, the retarded Bondi coordinates~\eqref{Bondi-coords} fit within this class of parameterizations with the identifications
\begin{equation} \label{Bondi-ADM}
    N^a = \delta^a{}_u \, , \qquad h_{ab} \text{d}x^a \text{d}x^b = - \text{d}u^2 + r^2 \gamma_{ij} \text{d}x^i \text{d}x^j \, .
\end{equation}
These imply the properties
\begin{equation} \label{Bondi-cov}
    N^a N_a = - 1 \, , \quad \partial_r N^a = 0 \, , \quad D_a N^b = 0 \, , \quad N^a K_{ab} = 0 \, ,
\end{equation}
where boundary indices are raised and lowered via $h_{ab}$ and its inverse $h^{ab}$, $D_a$ is the Levi-Civita connection of $h_{ab}$, and $K_{ab}$ is the extrinsic curvature. In the current setup, the latter reads
\begin{equation}
    K_{ab} = \frac{1}{2}\, \partial_r h_{ab} \, ,
\end{equation}
and its only non-trivial components in retarded Bondi coordinates are $K_{ij} = r\, \gamma_{ij}$.
The relations \eqref{Bondi-cov} provide a characterization of our foliation that is invariant under diffeomorphisms on each time-like surface and that allows one to rewrite the metric in the equivalent form
\begin{equation} \label{FlatADM}
    \text{d}s^2 = 2 N_a(x^b) \text{d}x^a \text{d}r + h_{ab}(r,x^c) \text{d}x^a \text{d}x^b \, .
\end{equation}
When the conditions \eqref{Bondi-cov} hold, the non-vanishing Christoffel symbols are
\begin{equation}
    \Gamma^r{}_{ab} = - K_{ab} \, , \quad \Gamma^a{}_{rb} = K^a{}_b \, , \quad \Gamma^{a}{}_{bc} = \Gamma^{(h)\,a}{}_{bc} + N^a K_{bc} \, ,
\end{equation}
where $\Gamma^{(h)\,a}{}_{bc}$ are the connection coefficients associated to the induced metric $h_{ab}$. Notice that in the limit $r \to \infty$ this metric becomes degenerate and the shift vector $N^a$ spans its kernel, so that one recovers the two geometric objects that characterize the Carroll geometry of $\mathscr{I}^+$ (see, \emph{e.g.}, \cite{Duval:2014uva}).

As an illustration of the procedure, let us consider for simplicity the case $s=1$. In the coordinates \eqref{FlatADM}, eq.~\eqref{bilinear0} reads
\begin{align} \label{cov-theta}
    &\hat{\theta}^r[B,A] = - \sqrt{- h}\, \Big[ A^a \left(\partial_r - N^b D_b\right) B_a - (1-\lambda)\, K^b{}_a A^a B_b \nn \\
    &\quad + \lambda \, N^b A^a D_a  B_b - (1-\lambda)\, N^a A_a \left( K N^b - D^b \right) B_b \Big] \, ,
\end{align}
where $K = h^{ab} K_{ab}$ and we imposed the radial gauge condition~$A_r = 0$. The falloffs \eqref{falloffs} can instead be translated in 
\begin{equation} \label{ADMCB}
    A_a \sim \mathcal{O}(1) \, , \qquad N^a A_a \sim \mathcal{O}(r^{-1}) \, ,
\end{equation}
and the solution space takes again the form $A^\mu = \delta_\epsilon A^\mu + A^\mu_{\mathrm{rad}}$ as in eq.~\eqref{decomposition}. Moreover, the following relations hold,
\begin{equation} 
    \partial_r \delta_\epsilon A_a = 0 \, , \quad N^a D_a \delta_\epsilon A_b = 0 \, , \quad N^a \delta_\epsilon A_a = 0 \, ,
\end{equation}
and they allow to cancel the divergences in \eqref{cov-theta} by adding the counterterm action
\begin{equation} \label{SctADM}
    \begin{split}
    S_{\mathrm{ct}} = \frac{\lambda-1}{2} \int \text{d}^{D-1}x \sqrt{-h}\, \Big[ & K^b{}_a A^a A_b -N^a A_a D^b A_b\\
    & + N^a A_b D^b A_a \Big] \, .
    \end{split}
\end{equation}
Using \eqref{Bondi-ADM}, the latter coincides with eq.~\eqref{eq:Lcts=1} (up to an integration by parts of $\cD_i$), but it is now manifestly invariant under boundary diffeomorphisms.




\end{document}